\documentclass[useAMS,usenatbib]{mn2e}

\usepackage{lipsum} 
\usepackage{graphicx}
\usepackage{amssymb}
\usepackage{color}
\usepackage{hyperref}
\hypersetup{pdftex,  
  breaklinks=true,  
  colorlinks=true,
  urlcolor=blue,
  linkcolor=darkorange,
  citecolor=darkgreen,
  }

\definecolor{orange}{cmyk}{0,0.4,0.8,0.2}
\definecolor{darkorange}{rgb}{.71,0.21,0.01}
\definecolor{darkgreen}{rgb}{.12,.54,.11}

\title[Mid-Infrared Period--Luminosity Relations of RR Lyrae Stars Derived from the AllWISE Data Release]{Mid-Infrared Period--Luminosity Relations of RR Lyrae Stars Derived from the AllWISE Data Release}
\author[C. R. Klein et al.]{C. R. Klein,$^{1}$\thanks{E-mail:
cklein@berkeley.edu} J. W. Richards,$^{2}$ N. R. Butler$^{3}$ and J. S. Bloom$^{1}$\\
$^{1}$Astronomy Department, University of California, Berkeley, CA 94720, USA\\
$^{2}$Department of Mathematics, San Francisco State University, San Francisco, CA 94132, USA\\
$^{3}$School of Earth and Space Exploration, Arizona State University, Tempe, AZ 85287, USA}
\begin{document}

\date{Accepted 2014 February 14. Received 2014 February 13; in original form 2014 January 30.}

\pagerange{\pageref{firstpage}--\pageref{lastpage}} \pubyear{2013}

\maketitle

\label{firstpage}

\begin{abstract}

We use photometry from the recent AllWISE Data Release of the {\it Wide-field Infrared Survey Explorer} ({\it \hspace{-1.5pt}WISE}\,) of 129 calibration stars, combined with prior distances obtained from the established $M_V$--[Fe/H] relation and {\it Hubble Space Telescope} trigonometric parallax, to derive mid-infrared period--luminosity relations for RR Lyrae pulsating variable stars. We derive relations in the $W1$, $W2$, and $W3$ wavebands (3.4, 4.6, and 12 $\mu$m, respectively), and for each of the two main RR Lyrae sub-types (RRab and RRc). We report an error on the period--luminosity relation slope for RRab stars of 0.2. We also fit posterior distances for the calibration catalog and find a median fractional distance error of 0.8 per cent.

\end{abstract}

\begin{keywords}
stars: variables: RR Lyrae -- stars: distances -- infrared: stars -- methods: statistical.
\end{keywords}

\section{Introduction}

As calibratable standard candles, RR Lyrae pulsating variable stars provide a means to obtain highly precise (1-2 per cent error) distance measurements within the range of 1-100 kpc. These old (age $\gtrsim10\times10^9$ yr) Population II objects permeate the Milky Way Halo and Bulge, and many reside in the Disc. \cite{1964ARA&amp;A...2...23P} and \cite{1995CAS....27.....S} both provide excellent reviews of this class of variable stars. We refer the reader to Section 5 of \cite{2006ARA&A..44...93S} for a review of optical ($V$-band) implementations of RR Lyrae stars as distance indicators.

It is well known that in optical wavebands the RR Lyrae period--luminosity relation is negligible (nearly no slope in the linear relation), but that at near- and mid-infrared wavebands the slope steepens and the relation is quite constrained. Figure 4 of \cite{2013ApJ...776..135M} illustrates this phenomenon, and indicates an expected asymptote in the RRab period--luminosity relation slope of around $-2.4$ to $-2.8$. Additionally, since much of the Milky Way RR Lyrae population resides behind significant interstellar dust, the markedly reduced extinction at infrared wavelengths [demonstrated in Figure 8 of \cite{2011ApJ...737...73F}] is often lower than photometric errors. Thus, at mid-infrared wavelengths, the need for dust extinction corrections to set mid-infrared luminosity vanishes.

In the present work we calibrate mid-infrared period--luminosity relations for RR Lyrae stars using the AllWISE Data Release of the {\it Wide-field Infrared Survey Explorer} ({\it WISE}\,) and NEOWISE missions \citep{2010AJ....140.1868W, 2011ApJ...731...53M}. Much of this work is a continuation of the methodology developed in \cite{2011ApJ...738..185K}, which performed a similar analysis using the {\it WISE} Preliminary Data Release. The earlier Preliminary Data Release included only the first 105 days of the {\it WISE} mission, covering 57 per cent of the sky. The AllWISE Data Release (made public 2013 November 13) covers the entire sky and combines all the {\it WISE} survey data from 2010 January 7 to 2011 February 1. By using this more comprehensive dataset the present analysis incorporates 129 RR Lyrae calibration stars, whereas  \cite{2011ApJ...738..185K} made use of only 76 RR Lyrae period--luminosity relation calibrators. Furthermore, the longer temporal baseline and improved photometric accuracy of the AllWISE Data Release provides for more accurate mean-flux magnitude measurements of the calibration stars.

A similar investigation into the mid-infrared period--luminosity relations of RR Lyrae stars is presented in \cite{2014arXiv1401.5523D}. The calibrations performed in \cite{2014arXiv1401.5523D} use RR Lyrae stars in 15 Galactic globular clusters, with distances extending beyond 15~kpc.\footnote{The calibration sample used in the present work, described in \S\ref{data_desc}, is comprised of nearby ($\leqslant 2.5$ kpc) stars with well-detected {\it WISE} light curves.} It is encouraging to note that the period--luminosity relation slopes reported in \cite{2014arXiv1401.5523D}, modulo a slight metallically dependance, agree so well with the calibrations published in the present work.

The applications for RR Lyrae stars as distance indicators are diverse and fundamental to many areas of astronomical inquiry. They are found throughout the Milky Way Halo and Bulge, pervade the Disc, and are still shining in neighboring dwarf galaxies.

This paper is outlined as follows. We present a brief description of the {\it WISE} and ancillary data in \S\ref{data_desc}. In \S\ref{methods} we review the employed light curve analysis methodology. In \S\ref{derivations} we present the derived period--luminosity relations, and in \S\ref{conclusions} we discuss the conclusions and future implications of this work.

\section[]{Data Description}\label{data_desc}
{\it WISE} provides imaging data in four mid-infrared wavebands: $W1$ centred at 3.4 $\mu$m, $W2$ centred at 4.6 $\mu$m, $W3$ centred at 12 $\mu$m, and $W4$ centred at 22 $\mu$m. Although the original {\it WISE} mission was designed for static science goals, the orbit and survey strategy of the {\it WISE} spacecraft (described in \cite{2010AJ....140.1868W}) are conducive to recovering light curves of periodic variables with periods $\lesssim 1.5$ day, which is well-matched to RR Lyrae variables.

In this analysis we make use of the most recent {\it WISE} data release, AllWISE. The AllWISE Data Release (made public 2013 November 13) combines the 4-Band Cryogenic Survey (main {\it WISE} mission covering the full sky 1.2 times from 2010 January 7 to 2010 August 6), the 3-Band Cryogenic survey (first three wavebands, 30 per cent of the sky from 2010 August 6 to 2010 September 29), and the NEOWISE post-cryogenic survey (first two wavebands, covering 70 per cent of the sky from 2010 September 20 to 2011 February 1). The individual photometry epochs were retrieved from the AllWISE Multiepoch Photometry Database.

As in \cite{2011ApJ...738..185K} we employ the catalog of 144 relatively local ($\leqslant 2.5$ kpc) RR Lyrae variables developed by \cite{1998AA...330..515F}. Fifteen of these stars are excluded from our present analysis because they were not well-detected by {\it WISE}. {\it WISE} photometry data with any quality flags were rejected for the period--luminosity relation fits (most common cause was confusion with neighbouring sources, in part due to the $\sim6.3$ arcsec PSF of {\it WISE}\,).

Also as in \cite{2011ApJ...738..185K}, {\it Hipparcos} photometry \citep{1997ESASP1200.....P} is transformed into $V$-band \citep{1998ApJ...508..844G}, corrected for dust extinction [using the line-of-sight extinction from \cite{1998ApJ...500..525S} and the $R$ factor from \cite{1975A&A....43..133S}], and combined with the \cite{postHipp..book} $M_V$--[Fe/H] relation to yield prior distance moduli, $\mu_{\rm prior}$. Precise trigonometric parallax angles for four of the stars (RRLyr, UVOct, XZCyg, and SUDra; all of the RRab subclass) have been previously measured with the {\it Hubble Space Telescope} and published in \cite{2011AJ....142..187B}.\footnote{RRc star RZCep also has an HST-measured parallax, but this star was rejected from the fit following the procedure described in \S\ref{methods}.} For these four stars the more precise, parallax-derived distance moduli are used in the period--luminosity relation fits. We note that the the distance moduli derived from the metallicity--luminosity relation for these four stars is in statistical agreement (within $2\sigma$) with the parallax-derived distances.

A significant difference between the present work and that of \cite{2011ApJ...738..185K} is that the two primary RR Lyrae subclasses (RRab and RRc stars) are now treated independently. Previously, in \cite{2011ApJ...738..185K} all the RR Lyrae stars were fit together, which [as noted by \cite{2013ApJ...776..135M}] is physically inappropriate because they follow different period--luminosity relations (RRab stars oscillate in the fundamental mode, whereas RRc stars do so in the first overtone). It is common practice to ``fundamentalize'' the periods of the RRc stars, as in \cite{2004ApJ...610..269D}, and use them to supplement the RRab period--luminosity fit. In the present analysis we instead treat the two subclasses independently, so as to calibrate both the RRab and RRc period--luminosity relations.

\section{Light Curve Analysis Methods}\label{methods}

The light curve analysis methods employed in the present work are an evolution of those described in \cite{2011ApJ...738..185K}. Mean-flux magnitudes are measured from harmonic model fits to the phase-folded {\it WISE} photometry. To more accurately assess the uncertainty associated with the measured mean-flux magnitude for each star, a parametric bootstrapping procedure was performed. The {\it WISE} photometry was resampled (assuming a normal distribution) and refit with a harmonic model 5,000 times to generate a distribution of mean-flux magnitude measurements. The standard deviation of this bootstrapped mean-flux magnitude distribution was taken to be the uncertainty used later in the period--luminosity relation fits.

The 5,000 harmonic models generated by the bootstrapping procedure were averaged to produce a mean harmonic model. This mean harmonic model yields a robust light curve amplitude. Furthermore, the standard deviation of the 5,000 harmonic models at each phase value provides a metric of how well the shape of the true light curve is recovered in the {\it WISE} photometry (if there is a lot of spread in the distribution of harmonic models, then the photometry is not accurate enough to reveal the shape of the true brightness oscillation). To improve the quality of the dataset used in the period--luminosity relation fits, any {\it WISE} light curve with a bootstrapped harmonic model maximum standard deviation larger than its robust amplitude measurement was excluded. This procedure serves to ensure that only stars with {\it WISE} light curves well-fit by the harmonic model (i.e., those exhibiting clear sinusoidal-like oscillation) are used in the period--luminosity relation fits. 

One final step before performing the period--luminosity relation derivations described in \S\ref{derivations} was to conduct a traditional least-squares linear regression for each relation independently. The resultant fitted zero points and slopes were incorporated into the full simultaneous Bayesian derivation as the starting values for the MCMC traces. Additionally, this procedure allowed for the identification and rejection of anomalous ($>2\sigma$) outliers.

The final dataset used in the period--luminosity relation fits is comprised of 104 RRab stars with $W1$ photometry, 104 RRab stars with $W2$ photometry, 66 RRab stars with $W3$ photometry, 19 RRc stars with $W1$ photometry, 19 RRc stars with $W2$ photometry, and 9 RRc stars with $W3$ photometry. Table \ref{tab:catalog} presents all of the apparent magnitude photometry used in the period--luminosity relation fits, as well as the prior distances and the resultant posterior distances calculated during the fitting.

%

\section[]{Period--Luminosity Relations}\label{derivations}

The present derivation of period--luminosity relations is very similar to the Bayesian approach first described in \cite{2011ApJ...738..185K} and later formalised in \cite{2012Ap&amp;SS.341...83K}. In brief, our statistical model of the period--luminosity relationship is
\begin{equation}
\label{eqn:general_PLR}
m_{ij}=\mu_i + M_{0,j} + \alpha_j \log_{10}\left(P_i/P_0\right) + \epsilon_{ij}
\end{equation}
where $m_{ij}$ is the observed apparent magnitude of the $i$th RR Lyrae star in the $j$th {\it WISE} waveband, $\mu_i$ is the distance modulus for the $i$th RR Lyrae star, $M_{0,j}$ is the absolute magnitude zero point for the $j$th waveband, $\alpha_j$ is the slope in the $j$th waveband, $P_i$ is the period of the $i$th RR Lyrae star in days, $P_0$ is a period normalization factor (we use $P_{0, {\rm RRab}} = 0.55$ day and $P_{0, {\rm RRc}} = 0.32$ day), and the $\epsilon_{ij}$ error terms are independent zero-mean Gaussian random deviates with variance $\left(\sigma \sigma_{m_{ij}}\right)^2$. The error terms describe the intrinsic scatter in $m_{ij}$ about the model, where $\sigma$ is a free parameter which is an unknown scale factor on the known measurement errors, $\sigma_{m_{ij}}$. We initialize $\sigma$ with a flat prior and find that its posterior distribution is approximately normally distributed with mean 1.42 (1.16) and standard deviation 0.08 (0.18) for the RRab (RRc) fit.

We fit the three linear relationships simultaneously using a Bayesian MCMC method. After the fit converges we draw 150,000 samples of the posterior model parameters. The posterior distributions are well-represented as Gaussian, so we report the traditional distribution mean and standard deviation in the zero points, slopes, and posterior distance moduli. The calibrated period--luminosity relations for RRab stars are:
\begin{eqnarray}
\label{eqn:RRab_PLRs}
M_{W1} = -0.495 \left(\pm 0.013\right) - 2.38 \left(\pm 0.20\right) \times \log \left(P/0.55\right) \\
M_{W2} = -0.490 \left(\pm 0.013\right) - 2.39 \left(\pm 0.20\right) \times \log \left(P/0.55\right) \\
M_{W3} = -0.537 \left(\pm 0.013\right) - 2.42 \left(\pm 0.20\right) \times \log \left(P/0.55\right)
\end{eqnarray}

The calibrated period--luminosity relations for RRc stars are:
\begin{eqnarray}
\label{eqn:RRc_PLRs}
M_{W1} = -0.231 \left(\pm 0.031\right) - 1.64 \left(\pm 0.62\right) \times \log \left(P/0.32\right) \\
M_{W2} = -0.216 \left(\pm 0.031\right) - 1.70 \left(\pm 0.62\right) \times \log \left(P/0.32\right) \\
M_{W3} = -0.232 \left(\pm 0.032\right) - 1.71 \left(\pm 0.65\right) \times \log \left(P/0.32\right) 
\end{eqnarray}

\begin{figure*}
	\centering
	\includegraphics{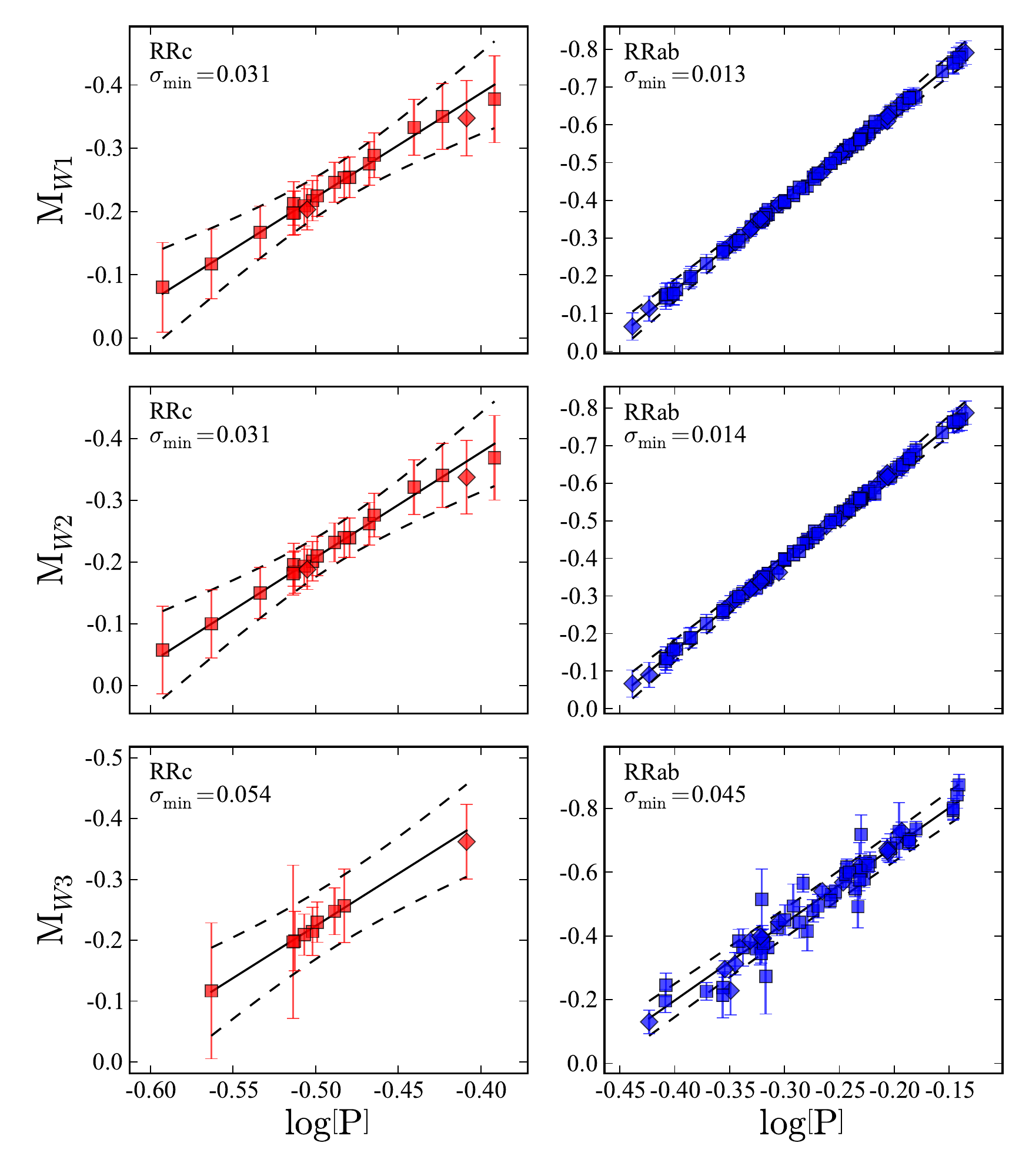}
	\caption{Period--Luminosity relations derived for each {\it WISE} waveband. Blazhko-affected stars, as identified via \url{http://www.univie.ac.at/tops/blazhko/Blazhkolist.html}, are denoted with diamonds, stars not known to exhibit the Blazhko effect are denoted with squares. The solid black line in each panel denotes the best-fit period--luminosity relation. The dashed lines indicate the 1-$\sigma$ prediction uncertainty for application of the best-fit period--luminosity relation to a new star with known period. See \S\ref{derivations} for further explanation, particularly with respect to the seemingly overestimated error bars.}
	\label{fig:PLRs}
\end{figure*}

The calibrated period--luminosity relations are plotted in Fig. \ref{fig:PLRs}. In each panel the solid black line denotes the best-fit period--luminosity relation, which corresponds to the above listed equations. The dashed lines indicate the 1-$\sigma$ prediction uncertainty for application of the best-fit period--luminosity relation to a new star with known period. The plotted points are the predictions for each star produced from jackknife fits (produced by withholding that star from input to a new fit, then using that new fit to predict the ``jackkniffed'' star's absolute magnitude). The absolute magnitude error on each datapoint is dominated by the prediction error of the fit (which is why it generally tracks with the 1-$\sigma$ prediction uncertainty envelope shown in dashed black lines). Note that these error bars do not correspond with the scatter of the data about the best fit line, nor should there be such a theoretical expectation. The minimum 1-$\sigma$ prediction uncertainty is given in the upper left of each panel. Blazhko-affected stars are indicated with diamond symbols, but they were not found to deviate from the fits. This is likely because the amplitude modulation at mid-infrared wavelengths is significantly reduced, but further study is needed to investigate this issue.

\begin{figure}
	\centering
	\includegraphics{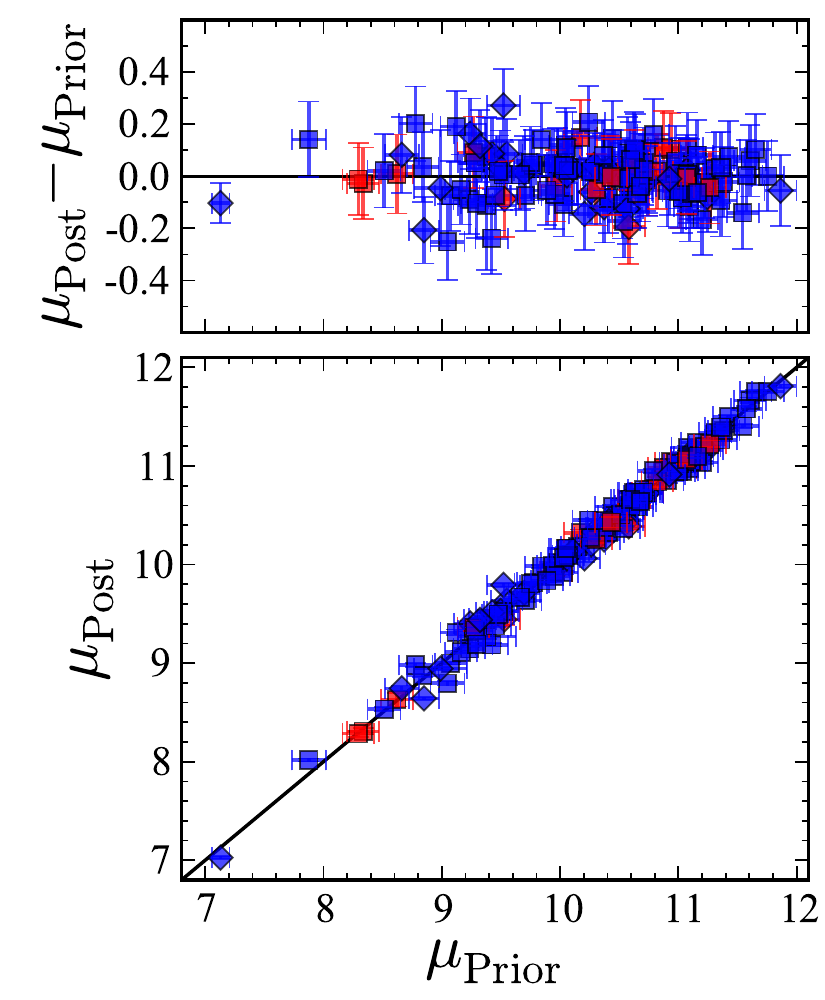}
	\caption{Prior vs posterior distance moduli. RRab stars are in blue, RRc stars in red. Blazhko-affected stars are denoted with diamonds, stars not known to exhibit the Blazhko effect are denoted with squares.}
	\label{fig:mu-mu}
\end{figure}

As a check on the fitted period--luminosity relations we compare the prior distance moduli, $\mu_{\rm prior}$, with the posterior distance moduli that were produced during the fitting, $\mu_{\rm post}$. Fig. \ref{fig:mu-mu} confirms that there is no obvious discrepancy between the distributions of the prior and posterior distances. The residual panel demonstrates that the posterior distances are in very good statistical agreement with the prior distances. The error bars here, in contrast with the prediction uncertainty error bars used in Figure \ref{fig:PLRs}, are derived from the $\mu_{\rm prior}$ and $\mu_{\rm post}$ distributions and should correspond with the scatter observed in the residual panel. In fact, 112 of 129 (87 per cent) of the residual data points are within one error bar length of zero, indicating that the errors are slightly overestimated.

The prediction uncertainties illustrated in Figure \ref{fig:PLRs} are intrinsic to the waveband-specific period--luminosity relations. The posterior distances produced through the simultaneous Bayesian linear regression, presented in Table \ref{tab:catalog} and plotted in Fig. \ref{fig:mu-mu}, are improved beyond the single-waveband prediction by using all three wavebands. This is representative of the advantages of employing a simultaneous fitting method. 

\section[]{Conclusions}\label{conclusions}

We have presented derivations of the period--luminosity relations at 3.4, 4.6, and 12 $\mu$m (first three wavebands of {\it WISE}\,) using AllWISE photometry for 129 calibrating stars. The employed Bayesian simultaneous linear regression fitting method yielded improved distances for these 129 calibrators with a median fractional error of 0.8 per cent.

Although the presented relations are intrinsic to the {\it WISE} photometric system, we expect that similarly well-constrained mid-infrared period--luminosity relations particular to the Spitzer Space Telescope or, eventually, the James Webb Space Telescope, can be constructed following the same methodology. Translating the {\it WISE} relations into other instrumental photometric systems is possible, but introduces significant systematic uncertainty.

The {\it WISE} spacecraft has recently been reactivated to observe with its $W1$ and $W2$ wavebands for three more years (2014$-$2016). This should allow for continued observations of a few thousand nearby RR Lyrae stars (within about 6~kpc), most of which have yet to be discovered and classified. The tightly-constrained mid-infrared period--luminosity relations will enable these stars to serve as very well-localized ``test particles'' in the Galactic Disc and Halo.

\begin{table*}
 \centering
 \begin{minipage}{177 mm}
  \caption{First few rows of the catalog of calibration RR Lyrae stars. The full table can be accessed as a text file in the publication's on-line data.}
  \begin{tabular}{lllrrrrrrrrrr}
  \hline
Name	& Type	& Blazhko		& Period	& \multicolumn{3}{l}{{\it WISE} Apparent Magnitudes}	&\multicolumn{3}{l}{Model Amplitudes}	& Prior Distance			& \multicolumn{2}{l}{Posterior Distance}\\
		&		& Affected?	& (day)	& $W1$ & $W2$ & $W3$ 						 	&$W1$	& $W2$	& $W3$			& $\mu_{\rm prior}$ 			& $\mu_{\rm post}$	& $d$ (pc) \\
		& 	 	&			&		& $\sigma_{W1}$ & $\sigma_{W2}$ & $\sigma_{W3}$	&		&		&				& $\sigma_{\mu_{\rm prior}}$	& $\sigma_{\mu_{\rm post}}$ & $\sigma_d$ (pc)\\ 
 \hline

AACMi 	& RRab 	& False		& 0.4763	& 10.238  	& 10.249	& 10.222	& 0.286	& 0.260	& 0.113	& 10.444 	& 10.587  & 1310 \\
          	&       	&			& 		& 0.006 	& 0.005	& 0.054	& 		& 		& 		& 0.146 	& 0.017 	& 10.3  \\

ABUMa 	& RRab 	& False		& 0.5996	& 9.570	& 9.592	& 9.530	& 0.173	& 0.136	& 0.070	& 10.044	& 10.163 	& 1078 \\
	 	& 	 	&			&		& 0.005	& 0.004 	& 0.026 	& 		& 		& 		& 0.141	& 0.016	& 8.2 \\

AEBoo	& RRc	& False		& 0.3149	& 9.713	& 9.721	& 9.665	& 0.099	& 0.099	& 0.083	& 9.965	& 9.928	& 967 \\
	 	& 	 	&			&		& 0.004	& 0.004	& 0.023	&		&		&		& 0.134	& 0.032	& 14.1 \\

AFVir	& RRab	& False		& 0.4837	& 10.698	& 10.712	& -----	& 0.266	& 0.203	& -----	& 11.093	& 11.063	& 1632 \\
	 	& 	 	&			&		& 0.006	& 0.009	& -----	&		& 		& 		& 0.134	& 0.017	& 12.6 \\

AMTuc	& RRc	& False		& 0.4058	& 10.569	& 10.578	& -----	& 0.104	& 0.103	& -----	& 11.019	& 10.970	& 1563 \\
		&		&			&		& 0.004	& 0.004	& -----	& 		&		&		& 0.133	& 0.068	& 49.3 \\


\hline
\label{tab:catalog}
\end{tabular}
\end{minipage}
\end{table*}

\section*{Acknowledgments}

The authors acknowledge the generous support of a CDI grant (\#0941742) from the National Science Foundation. J.S.B. and C.R.K. were also partially supported by grant NSF/AST-100991. This publication makes use of data products from the Wide-field Infrared Survey Explorer, which is a joint project of the University of California, Los Angeles, and the Jet Propulsion Laboratory/California Institute of Technology, funded by the National Aeronautics and Space Administration. This publication also makes use of data products from NEOWISE, which is a project of the Jet Propulsion Laboratory/California Institute of Technology, funded by the Planetary Science Division of the National Aeronautics and Space Administration. This research has made use of NASA's Astrophysics Data System.

\bibliographystyle{mn2e_alt}
\bibliography{Klein_refs}

\bsp

\label{lastpage}

\end{document}